\documentclass[a4paper,journal]{IEEEtran}
\usepackage{amsmath,amsfonts}
\usepackage{algorithmic}
\usepackage{array}
\usepackage[caption=false,font=normalsize,labelfont=sf,textfont=sf]{subfig}
\usepackage{graphicx}
\usepackage{cite}
\usepackage{url}
\usepackage{color}
\usepackage{hyperref} 
\usepackage[table]{xcolor}
\usepackage{blindtext}

\usepackage{dirtytalk}

\begin{document}

\title{Five Key Enablers for Communication during and after Disasters} 

\author{Mohammad Shehab, Mustafa Kishk, Maurilio Matracia, Mehdi Bennis, and Mohamed-Slim Alouini
\thanks{This work is supported by the KAUST Office of Sponsored Research under Award ORA-CRG2021-4695. The authors would like to thank the scientific illustrator Ana Bigio for her valuable contribution to Fig. 1.}
 
\thanks{Mohammad Shehab and Mohamed-Slim Alouini are with CEMSE Division, King Abdullah University of Science and Technology (KAUST), Thuwal 23955-6900, Saudi Arabia (email: mohammad.shehab@kaust.edu.sa, slim.alouini@kaust.edu.sa). 

Mustafa Kishk is with the Department of Electronic Engineering, Maynooth University, W23F2H6, Ireland. (email: mustafa.kishk@mu.ie). Maurilio Matracia is with Terna S.p.A., Palermo, Italy (email: matraciamaurilio@gmail.com). Mehdi Bennis is with the Centre for Wireless Communication, University of Oulu, Finland (email: mehdi.bennis@oulu.fi).
}
}

\maketitle

\begin{abstract}
Civilian communication during disasters such as earthquakes, floods, and military conflicts is crucial for saving lives. Nevertheless, several challenges exist during these circumstances such as the destruction of cellular communication and electricity infrastructure, lack of line of sight (LoS), and difficulty of localization under the rubble. In this article, we discuss key enablers that can boost communication during disasters, namely, satellite and aerial platforms, redundancy, silencing, and sustainable networks aided with wireless energy transfer (WET). The article also highlights how these solutions can be implemented in order to solve the failure of communication during disasters. Finally, it sheds light on unresolved challenges, as well as future research directions.

\end{abstract}


\section{Introduction} \label{sec:intro} 



 

During disasters, effective communication plays a crucial role in coordinating response efforts within the golden 72 hours and minimizing the impact on affected population. Timely and accurate information dissemination is essential for ensuring public safety, providing evacuation instructions, and coordinating emergency services. Various communication channels, including mass media, social media, instant messaging applications, and emergency alert systems are utilized to reach a diverse audience quickly. Clear and concise messaging helps alleviate panic and confusion, enabling individuals to make informed decisions about their safety. Additionally, communication facilitates the coordination of resources and assistance, allowing emergency responders to prioritize areas in need. Community engagement and two-way communication are also vital, as they enable authorities to gather valuable information from residents and address their concerns. The spread of internet connectivity, mobile applications and social media enhances the efficiency of communication strategies during disasters, ultimately contributing to more effective disaster management and response efforts \cite{survey}. The integration of  localization and sensing mobile technologies further facilitates the search and rescue operations during disasters. 

\begin{figure}[t!]
    \centering  \includegraphics[width=1\columnwidth]{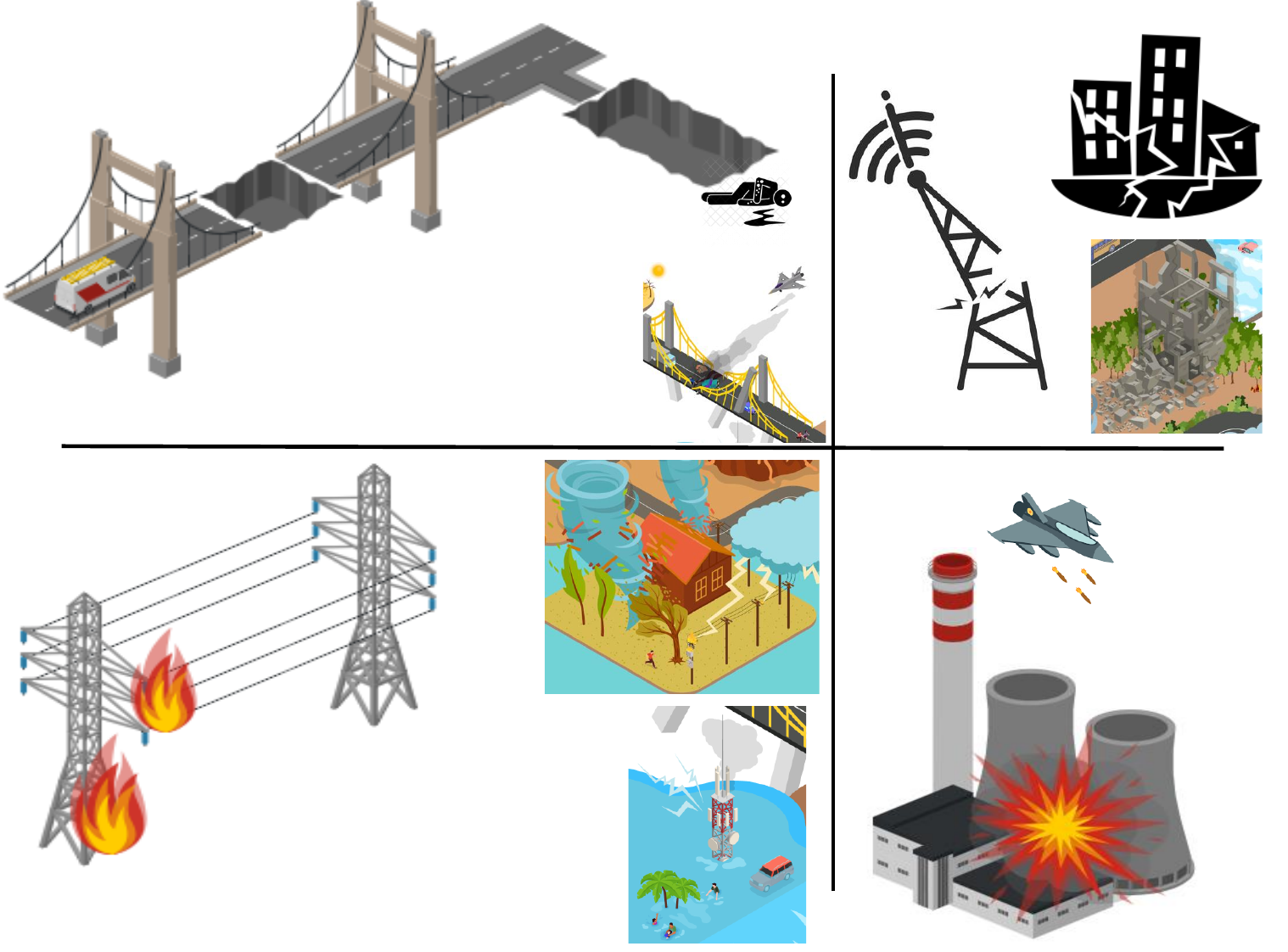} 
    \caption{Examples of infra-structure affected by disasters }
    \vspace{0mm}
    \label{disasters}
\end{figure}

As shown in Fig. \ref{disasters}, the destruction of communication infra-structure such as base stations (BSs) and electricity lines during disasters create lots of challenges. For instance, in areas affected by war-zones and earthquakes, cellular communication infrastructure and power cables could be damaged. This may effectively turn-off BSs and servers causing deterioration in coverage and in addition, reduce the ability of human users to charge their own mobile devices. Furthermore, among the challenges that face communication during disasters is the lack of LoS for communication and localization of victims under the rubble.

\begin{figure*}[t!]
    \centering    \includegraphics[width=1.68\columnwidth]{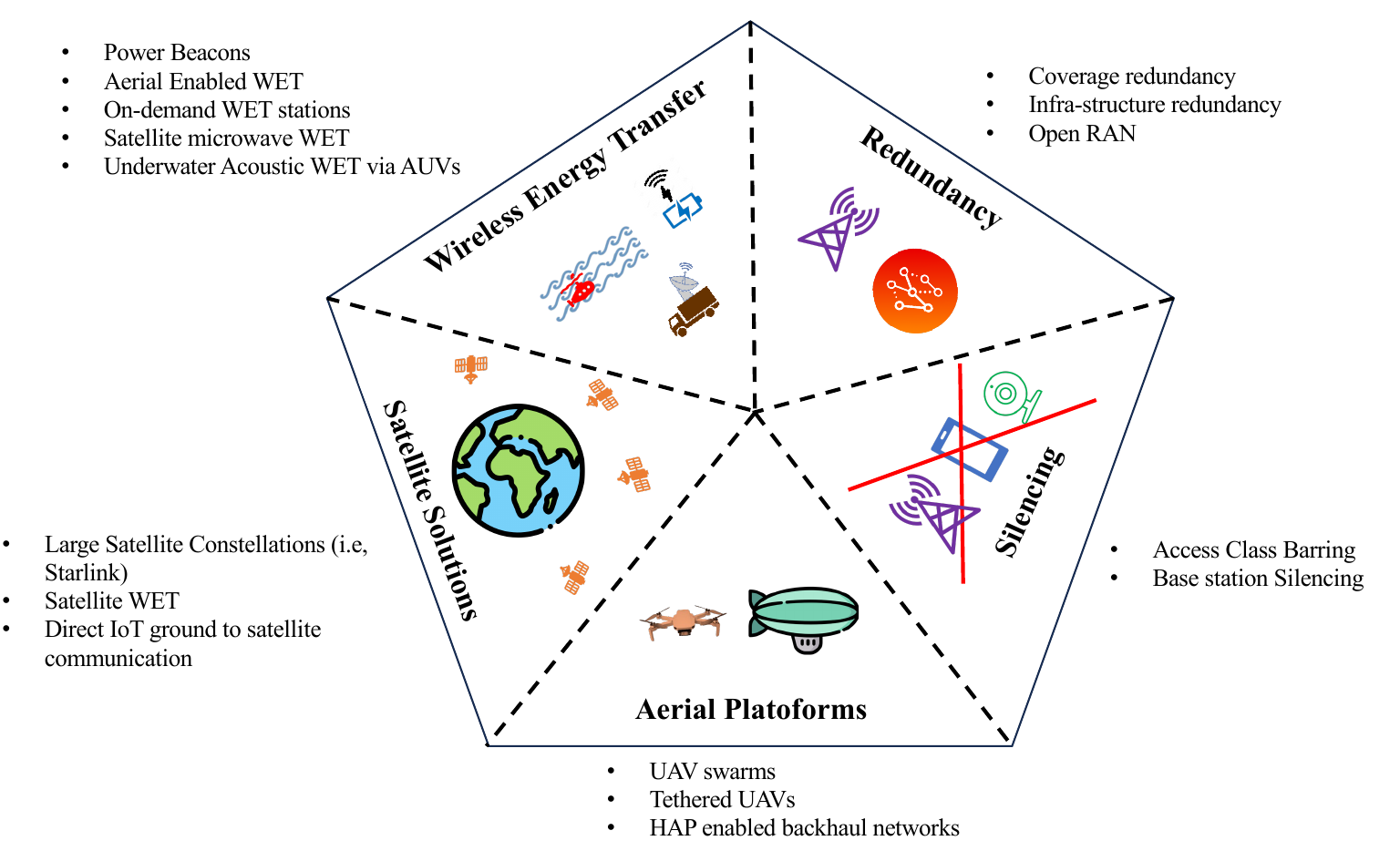} 
    \caption{An illustration of the proposed five key enablers for communication during disasters.}
    \label{Enablers}
\end{figure*}

With the purpose of connecting the unconnected during disasters, in this paper, we propose some enablers and discuss how to integrate these enablers to solve the above problems. The proposed solutions are shown in Fig. \ref{Enablers} as follows: aerial platforms \cite{aerial}, WET, satellite solutions, redundancy, and silencing \cite{silencing}. In the rest of the paper, we dive into the sea of these solutions illustrating how they can be applied to face the challenges that appear in Fig. \ref{disasters}. We also illustrate specific exemplary scenarios for satellite WET and fast communication recovery via BSs silencing.

\subsection{Selected Literature} \label{sec:SOTA}

Despite the importance and sensitivity of this topic, few research works have been tackling it. For instance, the authors of \cite{Kamran} introduced an energy harvesting approach for device to device relays that use harvested energy for communication, integrating clustering techniques to maintain services when cellular infrastructure is compromised. They demonstrated significant improvements in coverage, energy efficiency, and emergency resilience during disasters. Meanwhile, the work in \cite{survey} presented a rich survey on post-disaster communication including interesting insights on aerial platforms, localization, and routing.

\subsection{Outline}
In this work, we propose five key enablers for communication during disasters. First, in Section \ref{sec:aerial}, we illustrate the role of aerial platforms in disasters. Second, Section \ref{sec:WET} depicts different ways of WET that could assist in providing energy for devices in disaster areas affected by energy cut-offs. After that, Section \ref{sec:satellites} highlights the potential of large satellite constellations, satellite WET, and device to satellite communication in disaster scenarios. Section \ref{sec:redundancy} discusses network redundancy and O-RAN, while Section \ref{sec:silencing} tackles fast communication recovery via silencing approaches. The open challenges of applying these enablers are highlighted at then end followed by a summarized conclusion.

\section{Aerial platform} \label{sec:aerial}
\begin{figure*}[t!]
    \centering    \includegraphics[width=1.55\columnwidth]{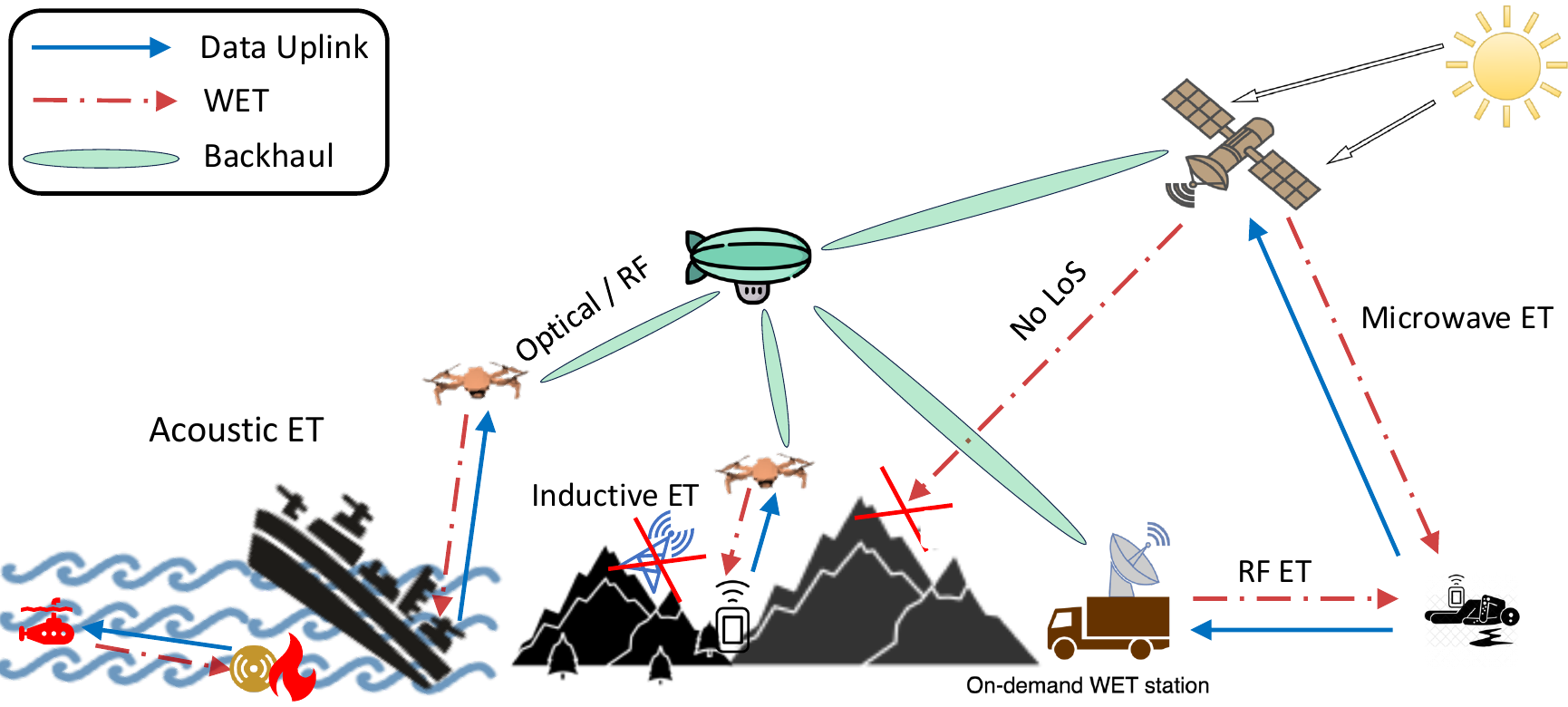} 
    \caption{Aerial platforms and WET solutions during disaster scenarios.}
    
    \label{Aerial_WET}
\end{figure*}

Unmanned aerial vehicles (UAVs) have revolutionized search and rescue operations in post-disaster scenarios by providing rapid, efficient, and comprehensive aerial assessments of affected areas. Equipped with high-resolution cameras, LiDAR, thermal and low frequency through wall sensing technologies, and GPS, UAVs can locate survivors, assess damage, and identify hazardous zones that might be inaccessible for human rescuers. As shown in Fig. \ref{Aerial_WET}, their ability to maneuver over debris and hard-to-reach locations allows for real-time data collection and situational awareness, significantly enhancing the speed and accuracy of rescue missions. UAV swarms can also deliver essential supplies such as medical kits, water, and food to stranded individuals, bridging critical gaps in the immediate aftermath of a disaster. However, this requires a high level of coordination in order to operate efficiently and avoid collision. This coordination can be achieved via high altitude platforms (HAPs), which can serve as a backhaul for the UAV swarmn etwork as shown in Fig. \ref{Aerial_WET}. Both UAVs and HAPs can act as flying BSs to provide coverage extension for areas affected by disasters.  In general, HAPs are more suitable for enhancing the coverage probability for large disasters, while UAVs provide more flexibility in small disasters areas \cite{survey}. Moreover, HAPs can serve as efficient relay of information between satellites and terrestrial devices to counter the satellite-to-ground path loss. 

An innovative aspect of UAV technology in search and rescue operations is the use of wireless signals to locate victims. A UAV network can communicate with the mobile devices of impacted individuals, using the exchange of signals or received signal strength to pinpoint or triangulate their exact positions. Among a plethora of sensing modalities, wireless (RF) sensors have several advantages as they can see through walls \cite{hackathon} and rubble by detecting reflected waves. Likewise, sensing and localization can be enabled by transmitting microwave or RF signals through walls to look for reflection changes, an approach deployed by NASA, using its FINDER program, during the Nepalese 7.8 magnitude earthquake in 2015. Such capabilities are particularly valuable in situations where visual identification is challenging, such as in dense urban environments or heavily forested areas. The continuous exchange of wireless signals ensures that rescuers can maintain an accurate and up-to-date understanding of victim location and vital data, facilitating faster and more targeted rescue efforts.

In addition to localization, UAVs can overcome the challenge of communicating with victims whose mobile devices are out of battery by utilizing WET. This approach allows UAVs to remotely recharge mobile devices by emitting targeted energy beams. Once the phone is powered on, the UAV can establish a wireless connection and receive essential data from the victim. This dual functionality not only enhances the effectiveness of search and rescue missions, but also significantly increases the chances of finding survivors who might otherwise remain undetected. 

The integration of advanced methods such as machine learning (ML) and sensor fusion with UAVs has further augmented their capabilities in object detection and search and rescue operations. For instance, the Hamilton-Jacobi-Bellman (HJB) and Fokker-Planck-Kolmogorov (FPK) equations necessary for UAV swarm control require powerful processing to be solved. In this context, the work in \cite{hamid} utilized mean-field games and federated learning to control a massive population of UAVs with limited inter-UAV communication in a way that efficiently plans their trajectories and avoid collisions among the UAVs. Their algorithm included a deep learning based function approximator that approximates the solutions of HJB and FPK equations, making them suitable for onboard UAV hardware. In another context, ML algorithms can analyze UAV footage to identify patterns and detect signs of human life, while ML can improve these algorithms over time based on new data. This technology enables UAVs to conduct autonomous searches and even predict the most probable locations of survivors based on historical disaster data and environmental factors. Another example of the merits of ML-aided UAV swarms appears in \cite{MARL}, where the authors presented low complexity trajectory planning solutions for a swarm of UAV collecting information from grounded IoT devices. They applied different multi-agent reinforcement learning (MARL) schemes such as cooperative and partially cooperative MARL, which could be extended to localization and sensing scenarios. In principal, the adoption of ML-aided UAVs not only enhances the effectiveness of rescue efforts but also reduces risks to human rescuers, allowing them to focus on the most critical tasks, while UAVs handle preliminary assessments and monitoring.

Despite the numerous advantages of UAVs in search and rescue missions, one significant challenge is their limited flight duration. Most commercially available UAVs have a battery life that ranges from 20 to 60 minutes, which can severely constrain their operational time in critical situations. This limitation necessitates frequent returns to base for battery replacement or recharging, which interrupts search patterns and reduces the overall efficiency of the mission. In large-scale disaster areas, where the terrain to cover is vast and the urgency is high, these interruptions can lead to delays in locating survivors and delivering aid, potentially costing valuable time and lives.

To mitigate the impact of limited flight duration, several strategies are being explored. One approach is the deployment of multiple UAVs in a coordinated fleet, ensuring that as one UAV returns for recharging, another can take its place, thus maintaining continuous aerial coverage. Tethered UAVs are also a very attractive solution to counter the issue of limited flight time due to UAV battery constraints. However, the flight range will be limited by the cable length.  Additionally, advancements in battery technology and energy-efficient designs are being pursued to extend the operational time of UAVs. Solar-powered UAVs and those equipped with rapid charging systems are also being developed to enhance endurance. Furthermore, integrating UAVs with ground-based charging depots distributed throughout the disaster area can provide a more seamless solution, allowing UAVs to recharge more frequently without long returns to a central base. In the context of WET, using laser-powered UAVs is a very promising solution currently being developed with the potential of significantly extending the flight duration by powering the UAVs using laser beams directed from ground depots, other UAVs or HAPs. This idea has already been investigated in \cite{amr} among many other references.

\section{Wireless Energy Transfer} \label{sec:WET}
WET plays an important role in the case of power outage within the area of disaster. Low power wireless and electronic devices can be remotely charged via terrestrial power beacons, aerial platforms or satellites. For instance, a victim's phone running out of battery could be wirelessly charged, which enables this victim to call emergency, receive and send information on their injury, or assist in localization of the victim under the rubble. Moreover, WET may provide the opportunity for waking up sensors and cameras to call back what happened in specific location where there is shortage of power. Several WET technologies exist to address these kind of scenarios. Table \ref{WET Technologies} as well as Fig. \ref{Aerial_WET} elucidate WET technologies suitable for different environments and the suggested WET provider platforms for each case. For example, optical WET is convenient for charging at long distances in the presence of LoS. Meanwhile, acoustic waves via autonomous underwater vehicles (AUVs) are efficient, safe, and suitable for long distance underwater environments. More commonly, capacitive and inductive technologies are applied for short distances and low energy devices if the charger can get closer to the target devices \cite{WET1}. Hybrid methods that combine several modes of WET also exist. It is clear from the table and the figure that in terrestrial setups when LoS exists, providing WET for disaster areas via aerial and satellite platforms is a promising solution that does not require any pre-existing infrastructure.

WET via satellites and aerial platforms revolutionizes the transmission to hard-to-reach areas. Satellites equipped with solar panels capture sunlight in space, converting it into electrical energy. This energy is then transmitted wirelessly and harvested by receivers on Earth, eliminating the need for traditional cables or infrastructure \cite{sat_wet}. This innovation holds immense potential for disaster areas lacking access to electricity and communication infrastructure. As the technology continues to evolve, it promises to be a reliable solution in such situations.

Performing WET via satellites involves several key steps and technologies. First, the satellite must be equipped with solar panels to capture solar energy and convert it into electrical power. Next, this electrical energy is converted into a specific frequency, such as microwave or laser beams, for transmission. These beams are directed towards the target devices on earth, which should equipped with specialized antennas designed to capture and convert the transmitted energy back into electricity.

WET using satellites faces several significant challenges that need to be addressed for its widespread implementation. One major challenge is the efficiency of energy conversion and transmission. Despite advancements in technology, a considerable amount of energy is lost during the conversion and transmission process, reducing the overall efficiency of the system. Additionally, maintaining accurate alignment between the satellite and receiving devices on Earth poses a challenge, especially in dynamic environmental conditions or when dealing with moving targets. Another critical consideration is the potential for electromagnetic interference with other satellite communication systems or terrestrial infrastructure. Safeguarding against interference requires careful frequency management and coordination. Moreover, ensuring the safety of both the satellite-based energy transmission and the receiving end devices on earth is of paramount importance. Lastly, the high initial investment and infrastructure costs associated with deploying satellite-based energy transfer systems present economic barriers to widespread adoption. Addressing these challenges will be essential for realizing the full potential of WET via satellites as a sustainable and efficient energy solution in disaster areas.

\begin{table}[!t]
\centering
    \caption{WET technologies and providers for different environments}
	\label{WET Technologies}
\begin{tabular}{| m{2.2cm} | m{2.5cm}| m{2.2cm} |} 
\hline 
\rowcolor{cyan} \vspace{0.9mm} Environment \vspace{0.9mm} &\vspace{0.9mm} Technology \vspace{0.9mm} & \vspace{0.9mm} WET Provider \vspace{0.9mm} \\ 
\hline
$\bullet$ Underwater & $\bullet$  Acoustic & \vspace{0.5mm} $\bullet$ HAPs

$\bullet$ AUVs

$\bullet$ Buoy stations\\
\hline                              
$\bullet$ Mountains 

$\bullet$ Rubble 

$\bullet$ No LoS & $\bullet$  Inductive/capacitive

$\bullet$  RF  & $\bullet$  UAVs\\
\hline                              
$\bullet$ Clear Terrain 

$\bullet$ LoS & $\bullet$ Optical (e.g, laser)

$\bullet$ Microwave 

$\bullet$ RF   & \vspace{1mm} $\bullet$    Satellites \vspace{0.2mm}

$\bullet$ HAPs

$\bullet$ UAVs  \vspace{-2mm} 
\begin{flushleft} $\bullet$ On-demand ground stations  \end{flushleft} \\ 
\hline                              
\end{tabular} 
\end{table}

In order to learn more technical details about the potential of satellite communication and satellite WET in times of disasters, we jump to the next section.

\section{Satellites} \label{sec:satellites}
Due to their inherent broadcast capabilities, global availability, wide coverage, satellites can play important an important role in times of disasters. During natural disasters or emergencies, when terrestrial infrastructure may be damaged or unavailable, satellites can maintain communication links. They are essential for coordinating rescue operations and providing aid in affected areas. Large satellite constellations offer a promising solution for connectivity in areas affected by disasters. Starlink \cite{star2} is originally a satellite internet constellation project developed by SpaceX and aims to provide high-speed, low-latency broadband internet access globally, particularly in underserved and remote areas. Launched in 2015, the initiative seeks to revolutionize internet connectivity by deploying a network of small satellites in low Earth orbit. With thousands of satellites already in orbit and plans for further expansion, Starlink offers potential to provide coverage for civilians affected by catastrophes and wars. This has been already been implemented during the recent war in Ukraine and the aftermath of the 2023 earthquake in Morocco \cite{hackathon}. Not only this, satellites also offer a globally available WET solution that can charge devices on earth from space.

\begin{figure}[t!]
    \centering    \includegraphics[width=1\columnwidth]{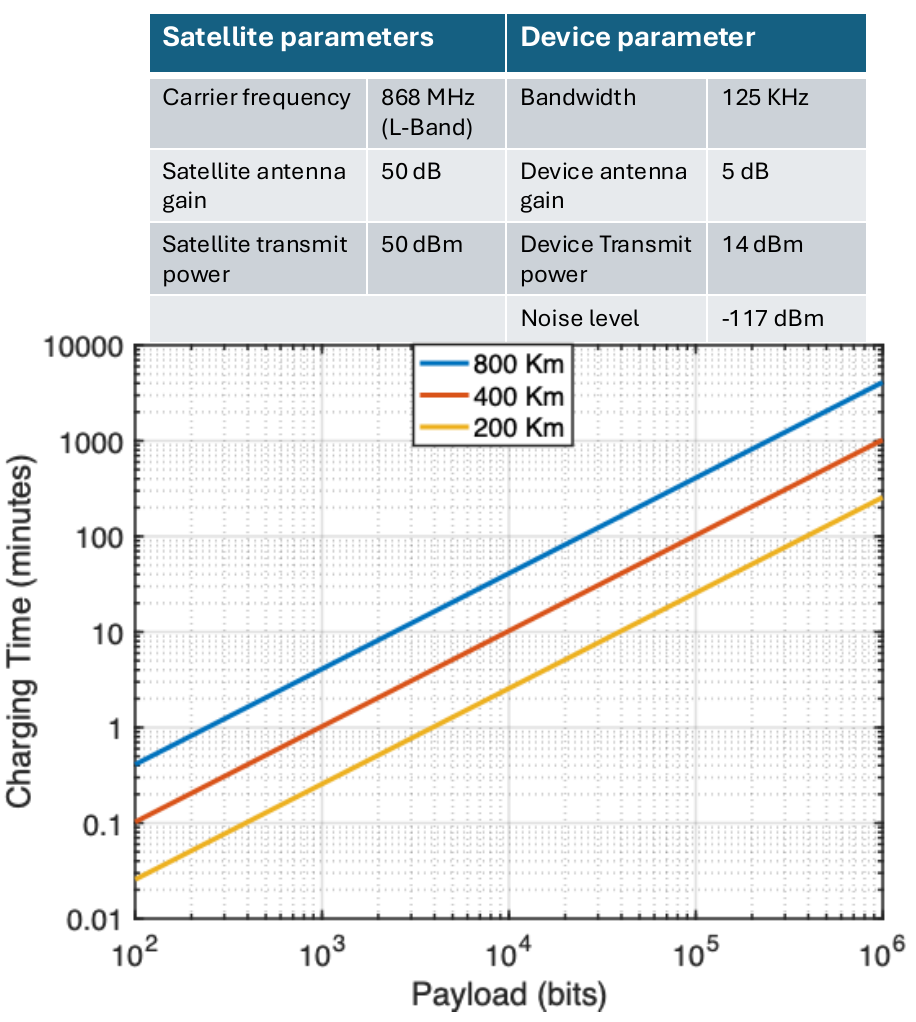} 
    \caption{Satellite charging time for different payloads}
    \label{satellite}
\end{figure}

In order to illustrate the idea of WET from space, we consider an EU 868 MHz ISM band LEO satellite to ground microwave link with transmit power of 50 dBm and transmit antenna gain of 50 dB. Simulation results depict that if the satellite height is 200 Km, a ground node can harvest an average of 3 nwatts of power from that satellite. Interestingly, over time, this harvested power level can mount up to a sufficient amount of energy to deliver few packets of data from the ground device to the nearest BS. To elaborate further, consider a scenario where a ground device collects energy from the satellite in order to transmit a specific payload to the nearest BS. Fig. \ref{satellite} depicts the charging time needed for the device to harvest enough energy to transmit this payload. For instance, from the figure, we can observe that, in order to transmit a small payload of 400 bits, the charging time from a satellite at 200 Km height is only 0.1 minute (i.e, 6 seconds). Furthermore, the device would require less than 1 minute of charging from a satellite at 200 or 400 Km in order to transmit up to 1 Kbits of information. However, the charging time required to transmit larger payloads such as 1 Mbits may rise to hours in that case.

In the same context, several works have shown the ability of grounded IoT devices to upload packets directly to satellites in times of disasters. These packets may include for instance, imaging or important information about signs of life. Interestingly, the work in \cite{Asad} depicted that for the same setup in Fig. \ref{satellite}, information can be transmitted directly to LEO satellite-based gateways, with high probability of successful packet reception (i.e, $\approx 0.9$) even in dense networks. The results in that work illustrated that adjusting the spreading factor according to the distance between the grounded device and the satellite plays an important role in improving the success probability. However, several factors such as dynamic beamforming and adaptive modulation and coding schemes are key ways for further enhancement.

\section{Redundancy and Disaster Anticipation} \label{sec:redundancy}

Redundant equipment refers to additional, backup systems or devices designed to ensure continued operation and reliability in case the primary equipment fails. The pre-installation of extra equipment to monitor and anticipate the occurrence of disasters or redundant infra-structure serving as a reserve for existing equipment may require extra costs. However, it can payoff at the time of disasters by facilitating the search and rescue operations and saving lives. This could specially be efficient in areas, where specific disasters such as earthquakes or floods are common to occur. Herein, we list some few examples of how redundancy might be implemented:

\begin{itemize}
    \item  Dual or Multiple Servers: Using multiple servers to host communication applications so that if one server fails, others can take over.

    \item Backup Power Supplies: Installing uninterruptible power supplies (UPS) or backup generators to keep communication equipment operational, if the main power supplies are damaged.

    \item Replicated Base stations: Deploying duplicate BSs at different locations to ensure continuity if a primary site is compromised during disaster.

    \item On-demand service providers (SP): Aerial, grounded or underwater on-demand BSs or SPs can be dynamically deployed around areas, where disasters are expected to occur. Those SPs may come in to the area of disaster swiftly in short time.   
\end{itemize}

Founded in 2018, Open RAN (O-RAN) \cite{ORAN} alliance is a global entity that includes many mobile network operators, vendors, manufacturers, and research organizations. It aims at defining specifications for O-RAN components that ensure openness and interoperability. The remote radio heads and base band units are disaggregated to radio units, distributed units and centralized units with open interfaces and open software that enable interoperability from multiple suppliers. Key components of O-RAN include cloudification, open and interoperable interfaces and intelligent network access. The flexibility offered by O-RAN may facilitate the deployment of missing or destroyed components from several sources during times of disasters. Not only this, O-RAN allows for installing redundant components from multiple vendors at minimum cost.

Another solution to enable fast recovery after a disaster is silencing, which can be instantly performed the case of absence of extra equipment as will be shown in the next section.

\section{Silencing} \label{sec:silencing}
\color{black} 

\begin{figure}[t!]
    \centering    \includegraphics[width=1\columnwidth]{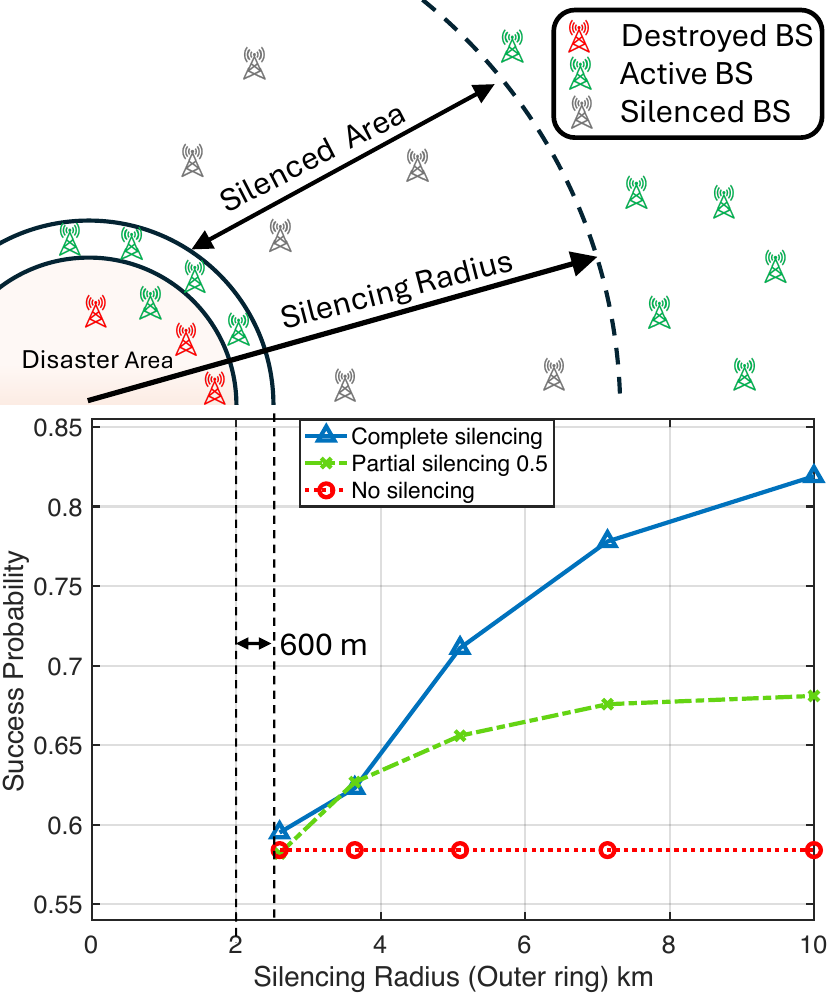} \vspace{0mm}
    \caption{Typical base station silencing setup in post-disaster scenarios: while the disaster area is characterized by non-functioning BS, the silencing area presents BSs which transmit power is suppressed or completely turned-off, and everywhere else the BSs are still functioning.}
    \vspace{0mm}
    \label{silencingSetup}
\end{figure}

BS silencing \cite{silencing,silencing2} and application access class barring (ACB) are rapid interference reduction solutions that can be adopted to boost communication in case of disasters. One typical setup for BS silencing is depicted in Fig.~\ref{silencingSetup}), where we apply stochastic geometry by assuming a Poisson point process of devices and BSs and a circular disaster affecting an area within 2 Km radius, where some of the BSs in this area are destroyed. BS silencing can be implemented by turning off some of the BSs located in specific areas outside to the disaster area. This would reduce the aggregate interference affecting the devices inside the disaster area and allow them to transmit their signal to the active BSs inside and around the disaster area with low the energy consumption, whenever it is necessary.

Herein, BSs are allowed to be active within a ring of 600 m outside the disaster and the BSs outside this ring upto a specific silecning radius are silenced or suppressed. At a threshold of -10 dB and assuming no silencing, the success probability of receiving signals from devices within the disaster area is about 0.58. The figure depicts that this success probability can be significantly enhanced upto 0.82 via complete silencing of the BSs withing the silencing area. Further adjustment can be done by optimizing the silencing area or the silencing factor, based on the topology of the post-disaster network. However, this technique affects the overall capacity of the network and hence has to be carefully weighed by selecting the proper density of BSs to silence (based on an accurate load analysis).

Note that, it might be important to allow some coverage outside the disaster area to account for preparations for rescue and coordination efforts within that area. An interesting alternative consists of adjusting the BSs' transmit power in order to optimize the trade-off between the reduction of aggregate interference experienced by the users inside the disaster area and the coverage experienced by the users inside the silencing area. In that case, the BSs would not really be completely silenced, but rather partially silenced. The figure depicts that this option would partially enhance the success probability to 0.68. Finally, in case of spectrum abundance, it would be beneficial to design effective spectrum allocation schemes so that the BSs that are supposed to be silenced can actually keep transmitting signals within a different frequency band, and hence, serve their respective users without interfering with the ones inside the disaster area.


Another form of silencing is the application ACB \cite{ACB}, where applications are categorized according to application specific congestion control (ACDC) categories. ACDC is an access control mechanism that classifies the user side applications according to their priority. In times of disasters, applications with the highest ACDC (e.g, localization and instant messaging applications) may have higher priority to access the network for more time. Meanwhile, the access time is significantly reduced (or the access is completely not allowed) for applications with low ACDC category such as gaming applications. To this end, the silencing of applications with lower ACDC would reduce the network congestion and give more bandwidth availability to the essential applications for search and rescue operations.


\section{Open Challenges} \label{sec:discussion}
Despite the recent advances and the solutions provided, there are still some serious challenges that require further investigation such as:

\begin{itemize}

\item Integration: The study in \cite{survey} illustrated the integration of satellites with HAPs and UAVs in disaster situations. However, the integration of the proposed solutions into the existing cellular world requires further investigation. For instance, more research and standardization efforts are required on how to coordinate and integrate UAVs and HAPs into the existing cellular networks. 

\item Hardware requirements: Delivering energy from space or HAPs seem to be very promising ideas. However, does the devices contain the necessary on-board hardware required to receive this power and harvest it? what would be the implementation cost? These questions pose a serious challenge that requires further focus on the hardware level.

\item Satellite WET: Most research on satellites focus on data transmission. Interestingly, this paper illustrated that satellite WET poses a great potential for further investigation in remote and disaster situations. However, proper planning of spectrum and scheduling mechanisms should be considered in order to mitigate interference with communication channels operating on the same frequencies. Moreover, the amount of harvested energy could be further enhanced by using more efficient and directive antenna, increasing the satellite transmit power, and adopting lower transmit frequencies.

\item Rubble channel modelling: Communication with people or devices stuck under the rubble is a great challenge due to the complex channel conditions, thus requiring proper channel modelling and adaptive schemes.

\item Political Challenges: In many occasions, there exist political challenges that hinder the implementation of the suggested solutions during military conflicts (e.g, 2024 Gaza war). Therefore, compulsory laws should be globally imposed to allow civilian communication enablers via neutral global entities such as the UN and the Red Cross. 

\end{itemize}

\section{End Line} \label{sec:discussion}

In this paper, we presented an overview on communication during disasters highlighting the main challenges and five key enablers. We discussed the role of aerial platforms and large satellite constellations in disasters. WET options in different disaster environments were highlighted including an illustration of novel scenario for Satellite to ground WET. We elucidated the importance of redundant infra-structure and O-RAN in disaster management. Furthermore, the paper presented BS silencing and ACB as attractive and fast methods for restoring connectivity in disaster areas. Finally, we shed light on the open challenges that face the implementation of the proposed enablers among which are rubble channel modeling, hardware requirements, and political challenges.

\bibliographystyle{IEEEtran}
\bibliography{main}

\section*{Biographies}
\footnotesize

\noindent\textbf{MOHAMMAD SHEHAB} is a postdoctoral Researcher with CEMSE Division, King Abdullah University of Science and Technology (KAUST), Saudi Arabia. Previously, he was with the University of Oulu, Finland, from which he obtained his doctoral degree in 2022. Prior to that, he obtained two M. Sc degrees from the Arab Academy and University of Oulu in 2014 and 2017, respectively and worked as TA at Alexandria University and the Arab Academy in Egypt from 2012-2015. Shehab was a visiting researcher to TU Dresden, Aalborg University, and the American University in Cairo during 2018, 2019, and 2022, respectively. His current research interests include but are not limited to Machine Learning, UAVs, communication in extreme environments, and Semantic communication. Among a list of other awards, Mohammad won the Nokia Foundation award consecutively in 2018 and 2019. \\

\noindent\textbf{MUSTAFA A. KISHK} is an Assistant Professor with Maynooth University, Kildare, Ireland. He received the B.Sc. and M.Sc. degrees in electrical engineering from Cairo University, Giza, Egypt, in 2013 and 2015, respectively, and the Ph.D. degree in electrical engineering from Virginia Tech, Blacksburg, VA, USA, in 2018. From 2019 to 2022, he was a Postdoctoral Research Fellow with the King Abdullah University of Science and Technology, Thuwal, Saudi Arabia. His research interests include UAV communications, satellite communications, and global connectivity for rural and remote areas. \\

\noindent\textbf{MAURILIO MATRACIA} is currently attending the \textit{Tyrrhenian Lab} program organized by \textit{Terna S.p.A.} in Palermo, Italy.
He got his Ph.D. Degree from KAUST in 2023. 
His experience includes serving as a reviewer for several IEEE journals and receiving prizes in competitions such as the \textit{SusTech 2021 student poster} and the \textit{ComSoc EMEA Region $-$ Internet for All}.
His main research interest is stochastic geometry, with a special focus on post-disaster and rural wireless networks.\\

\noindent\textbf{MEHDI BENNIS} is a full (tenured) Professor at the Centre for Wireless Communications, University of Oulu, Finland and head of the intelligent connectivity and  networks/systems group (ICON). His main research interests are in radio resource management, game theory and distributed AI in 5G/6G networks. He has published more than 300 research papers in international conferences, journals and book chapters. He has been the recipient of several prestigious awards including the 2015 Fred W. Ellersick Prize from the IEEE Communications Society, the 2016 Best Tutorial Prize from the IEEE Communications Society, the 2017 EURASIP Best paper Award for the Journal of Wireless Communications and Networks, the all-University of Oulu award for research, the 2019 IEEE ComSoc Radio Communications Committee Early Achievement Award and the 2020–2023 Clarviate Highly Cited Researcher by the Web of Science. \\

\noindent\textbf{MOHAMED-SLIM ALOUINI}  [S’94, M’98, SM’03, F’09] is a Distinguished Professor of Electrical Engineering at King Abdullah University of Science and Technology (KAUST), Saudi Arabia. He received the Ph.D. degree in Electrical Engineering from the California Institute of Technology (Caltech), USA, in 1998. Currently, he holds the UNESCO Chair in Education to Connect the Unconnected and leads the activities of the Communication Theory Lab. His current research interests include modeling, design, and performance analysis of wireless communication systems.

\end{document}